\documentstyle[aps,prb,epsfig]{revtex}

\begin{document}

\newcommand{\beq}{\begin{equation}}
\newcommand{\eeq}{\end{equation}}
\newcommand{\beqa}{\begin{eqnarray}}
\newcommand{\eeqa}{\end{eqnarray}}
\newcommand{\Eq}[1]{Eq.~(\ref{#1})}        
\newcommand{\Eqs}[1]{Eqs.~(\ref{#1})}      
\newcommand{\Sec}[1]{Sec.~(\ref{#1})}        
\newcommand{\Ref}[1]{Ref.~(\onlinecite{#1})}  
\newcommand{\pdag}{{\phantom{\dagger}}}
\newcommand{\lllzi}{\l - 2\varepsilon_i}
\newcommand{\lllzip}{\l -2\varepsilon_{i'}}
\newcommand{\lllzj}{\l - 2\varepsilon_j}
\newcommand{\lllmi}{\l_m - 2\varepsilon_i}
\newcommand{\E}{{\mathcal{E}}}               
\newcommand{\bra}{\langle 0|}
\newcommand{\ket}{|0\rangle}
\newcommand{\ds}{{\vphantom{\dagger}}}
\newcommand{\Rmi}{{\bf R_{mi}}}
\newcommand{\wmi}{{\bf w_{mi}}}
\newcommand{\Tm}{{\bf T_m}}
\newcommand{\bm}[1]{{\mbox{\boldmath$#1$\unboldmath}}}

\def\b{\beta}
\def\d{\delta}
\def\g{\gamma}
\def\a{\alpha}
\def\s{\sigma}
\def\t{\tau}
\def\l{\lambda}
\def\L{\Lambda}
\def\e{\epsilon}
\def\r{\rho}
\def\d{\delta}

\def\be{\begin{equation}}
\def\ee{\end{equation}}
\def\beq{\begin{eqnarray}}
\def\eeq{\end{eqnarray}}

\title{Algebraic Bethe Ansatz for a discrete-state BCS pairing model}

\author{J. von Delft${}^{1,2}$\footnote{e-mail: 
jan.vondelft@physik.uni-muenchen.de} and R. Poghossian${}^1$\footnote{
on leave of absence from the Yerevan Physics Institute, 
Armenia\\ e-mail: poghos@th.physik.uni-bonn.de}}
\address{$^{1}$Physikalisches Institut der Universit\"at Bonn,
 Nu{\ss}allee 12, D--53115 Bonn, Germany\\
$^{2}$Sektion Physik, 
Ludwig-Maximilians-Universit\"at M\"unchen, 80333 M\"unchen, Germany}
\date{June 20, 2001}
\maketitle
\begin{abstract}
  
We show in detail how Richardson's exact solution of a
discrete-state BCS (DBCS) model can be recovered as a special case
of an algebraic Bethe Ansatz solution of the inhomogeneous XXX
vertex model with twisted boundary conditions: by implementing the
twist using Sklyanin's ${\bf K}$-matrix construction and taking the
quasiclassical limit, one obtains a complete set of conserved
quantities, ${\bf H_i}$, from which the DBCS Hamiltonian can be
constructed as a second order polynomial.  The eigenvalues and
eigenstates of the ${\bf H_i}$ (which reduce to the Gaudin
Hamiltonians in the limit of infinitely strong coupling) are exactly
known in terms of a set of parameters determined by a set of
on-shell Bethe Ansatz equations, which reproduce Richardson's
equations for these parameters.  We thus clarify that the
integrability of the DBCS model is a special case of the
integrability of the twisted inhomogeneous XXX vertex model.
Furthermore, by considering the twisted inhomogeneous XXZ model
and/or choosing a generic polynomial of the $\bf H_i$s as
Hamiltonian, more general exactly solvable models can be
constructed.  -- To make the paper accessible to readers that are
not Bethe Ansatz experts, the introductory sections include a
self-contained review of those of its feature which are needed here.

\end{abstract}
\pacs{PACS numbers:}

\section{Introduction and Summary}

In a series of pioneering experiments in the mid-1990's, Ralph, Black and
Tinkham observed a spectroscopic gap indicative of pairing correlations in Al
nanograins \cite{RBT} that were so small that their electronic excitation
spectra were discrete. Their results inspired a growing number of theoretical
studies of superconducting pairing correlations in nanograins with a fixed
number electrons (see Refs.~\onlinecite{vD,vDR} for recent reviews). These
works are based on a model, to be called discrete-state BCS model (DBCS model)
below, described by a reduced BCS Hamiltonian for a discrete set of
doubly-degenerate energy levels, with a pairing interaction that scatters
pairs of electrons from one level to the next.  The DBCS model was solved
exactly by Richardson in a series of papers starting in 1963:\cite{Richardson}
he explicitly
constructed all eigenstates and eigenenergies of the DBCS Hamiltonian in terms
of a set of energy parameters whose values
are determined by (numerically) solving a set of
algebraic equations, to be called ``Richardson's equations''.
Though his work had, for a long
time, been overlooked by the condensed matter community, it has recently
received increasing attention in the context of studying pairing
correlations in nanoscale Al grains, where the existence of an exact solution
has turned out to be as useful as it had been unexpected.

The existence of an exact solution to a nontrivial model of course immediately
raises the question whether it is related  to any of the standard
ways of exactly solving solvable models.  The goal of this paper is to show
that this is indeed the case: \emph{Richardson's solution of the DBCS model is
a special case of an algebraic Bethe-Ansatz solution of the so-called
inhomogeneous XXX vertex model with twisted boundary conditions.}

This insight builds upon a series of recent observations regarding
exact properties of the DBCS model: In 1997, Cambiaggio, Rivas and
Saraceno \cite{CRS97} showed (though unaware of Richardson's work)
that the DBCS model was integrable, and explicitly constructed all the
constants of the motion [cf.  Eq.~(\ref{eq:constants-of-motion-DBCS})
below].  In 2000, Amico, Falci and Fazio \cite{AFF} realized that the
DBCS integrals of motion are in fact very similar to the integrals of
motion of the XXX Gaudin model [cf.~(\ref{gaudinham}) below],
differing from the latter only by an
additional $S_z$ term, and that Richardson's equations are very
similar to the so-called Gaudin equations, differing from the latter
only by an additional constant term.  Now, it has long been known
(see, e.g., chapter 13.2 of Gaudin's book\cite{gaudinbook})
that the Gaudin model can be derived
from the inhomogeneous XXX vertex (IXXX) model with periodic boundary
conditions, by taking the so-called quasiclassical limit, and that,
correspondingly, the Gaudin equations can be derived by taking the
quasiclassical limit of Bethe Ansatz equations of the IXXX model. Since
Richardson's Ansatz satisfies the Gaudin equations modified by the additional
constant term, Amico, Falci and Fazio\cite{AFF}
referred to Richardson's Ansatz as an ``off-shell
Bethe Ansatz'', i.e. an Ansatz \emph{not} satisfying the Bethe-equations of the
original XXX model, but of a modified version thereof.  (The off-shell Bethe
Ansatz was  originally introduced by Babujian and Flume in a context quite
different than finding eigenstates and eigenvalues of integral models, namely,
to solve Knizhnik-Zamolodchikov differential equations arising in conformal
Wess-Zumino models \cite{babujianflume}.)

In this paper, we address the following question: can one construct a
vertex model, integrable by the algebraic Bethe Ansatz (ABA), whose
quasiclassical limit \emph{directly} gives the DBCS model, in other
words, which is directly solved by a normal ``on-shell'' Bethe Ansatz?
The answer is positive: we show that the sought-after model is an IXXX
model with \emph{twisted} (instead of periodic) boundary conditions,
which we shall call the TIXXX model; its transfer matrix yields, in
the quasiclassical limit, a complete set of conserved quantities, ${\bf
  H_i}$, from which the DBCS Hamiltonian can be constructed as a
second order polynomial. Our emphasis on twisted boundary conditions
is the main difference between our work and that of references
\onlinecite{AFF,ADLO1}. We implement the twist using the boundary
${\bf K}$-matrix construction of Sklyanin, which he introduced while
developing his method of separation of
variables,\cite{sklyanin-1,sklyanin-2,sklyanin-3} an alternative (and in some
cases more powerful) way to the ABA for constructing wave functions.
In fact, Sklyanin himself mentioned in a side remark
in Ref.~\onlinecite{sklyanin-2} that the
quasiclassical limit of the IXXX model with twisted boundary
conditions (using a diagonal $K$-matrix) produces a modified version
of the Gaudin model (though he was not aware, at the time, of the
connection of the latter to the DBCS model).

We hope that our work fully clarifies the origin of the integrability of
the DBCS model by explicitly constructing the integrable TIXXX model from
which the latter can be derived. Moreover, by this construction we pave the
way for using the powerful algebraic Bethe Ansatz machinery to calculate
various quantities that have not yet been studied for the DBCS model. For
example, there has recently been great progress in using the ABA to calculate
matrix elements (or form factors) and correlation functions in vertex models,
which, by building upon our work, could now fruitfully be applied to the DBCS
model, too \cite{AO2001}.

Our work also suggests ways for constructing integrable generalizations of the
DBCS model, by considering other vertex models with twisted boundary
conditions.  In fact, one such generalization, which is Bethe-Ansatz solvable,
has recently been found independently by Amico, Di Lorenzo and Osterloh
\cite{ADLO1}. They showed that by a slight generalization of the integrals of
motion of the DBCS model, another integrable model is obtained.  We shall show
that the latter can be obtained by taking the quasiclassical limit of the
inhomogeneous XXZ vertex model with twisted boundary conditions (TIXXZ model),
in complete ananology to the derivation of the DBCS model from the TIXXX
model.

Another interesting direction in which our work could be pursued, is
to consider boundary conditions with \emph{nondiagonal} $K$-matrices.
These generally lead to models which are not solvable by the ABA.  
However, their eigenstates and eigenvalues can, in many
cases, nevertheless be found using Sklyanin's method of separation of
variables.

The paper is intended to be accessible also to readers that are not thoroughly
familiar with the details of the algebraic Bethe Ansatz; those of its features
which are needed here are therefore introduced and reviewed in pedagogical
detail.  The structure of the paper is as follows. In
Section~\ref{Richardson-Gaudin} we introduce the DBCS and Gaudin models,
recall how their integrals of motion are constructed, and give the equations
(Richardson's or Gaudin's) that have to be satisfied in order to obtain
eigenstates and eigenvectors. In Sections~\ref{ABA} and \ref{sec:eigenstates},
we give a self-contained introduction to the ABA method, as applied to the XXX
and XXZ vertex models; since both are special cases of the so-called 6-vertex
model, we shall actually begin by discussing the latter in full generality,
before specializing later on.  In Section~\ref{ABA} we explain how the
Yang-Baxter equations satisfied by the ${\bf R}$-matrices of local Boltzmann
weights lead to the exchange relations for the components of the Monodromy
matrix ${\cal T}$. Furthermore, we derive the fact that the transfer matrices
for different spectral parameters commute, which is the underlying reason for
the integrability of the model.  In Section~\ref{sec:eigenstates} we exploit
the exchange relations of the components of the Monodromy matrix to construct
the eigenstates and eigenvalues of the model.  In Section~\ref{sec:Sklyanin},
we explain how the results of Sections~\ref{ABA} and \ref{sec:eigenstates} can
be straightforwardly generalized to the case of twisted boundary conditions
using Sklyanin's ${\bf K}$-matrix. In Section~\ref{sec:quasiclassical}, we
take the quasiclassical limit of the TIXXZ model, and show that one thereby
recoveres a generalized version of the DBCS model. We also show that if one
specializes these results to the TIXXX model, one recovers the DBCS model.
Section~\ref{sec:conclusions} contains some brief conclusions and an outlook
for future applications of our results.

\section{The DBCS and Gaudin Models}
\label{Richardson-Gaudin}

The DBCS model that is commonly used \cite{vD,vDR} to describe superconducting
pairing correlations in nanoscale metallic grains is defined as follows:
one consideres a reduced BCS Hamiltonian,
\begin{eqnarray}
  \label{eq:mod-hamiltonian}
  H = \sum_{j, \sigma= \pm} 
\varepsilon_{j} c_{j \sigma}^\dagger c_{j \sigma}
  -g  \sum_{jj'}  c_{j +}^\dagger c_{j -}^\dagger 
c_{j' -} c_{j' +} \; ,
\end{eqnarray}
for electrons in a set of pairs of time-reversed single-particle states $|j,
\pm \rangle$ with energies $\varepsilon_j$, which are scattered pairwise from
level $j'$ to $j$, with interaction strength $g$.  Richardson managed to solve
this model exactly, for an arbitrary set of levels $\varepsilon_j$ (degenerate
levels are allowed, but are to be distinguished by distinct $j$-labels, i.e.\ 
they have $\varepsilon_i = \varepsilon_{j}$ for $i \neq j$): Since any level
occupied by only a single electron does not participate in and remains
``blocked''  to the pairscattering described by $H$, the labels
of all such single-occupied levels are good quantum numbers.  The eigenstates
$|\alpha \rangle$ and corresponding eigenenergies $\E_\alpha$ of $H$ thus have
the follwoing general form:
\begin{eqnarray}
|\alpha \rangle &= & \prod_{i \in B} c_{i \sigma_i}^\dagger |\Psi_P\rangle \;
,   \label{eq:generaleigenstate}  \qquad 
\E_{\alpha} = \E_P +  \sum_{i \in B}  \varepsilon_i \; .
\end{eqnarray}
Here $B$ is the set of singly-occupied, blocked levels, and $|\psi_P\rangle$
is an eigenstate, with eigenvalue $\E_P$ and containing
precisely $P$ pairs of electrons, of a Hamiltonian $H_U$ which
has precisely the same form as the $H$ of Eq.~(\ref{eq:mod-hamiltonian}),
except that now
the $j$-sums  are restricted to 
 run only over the set $U$ of all \emph{unblocked} or non-singly-occupied
levels.  It is now convenient to introduce the pseudospin variables
\begin{eqnarray}
  \label{eq:pseudospin}
&  S_j^z = {1 \over 2} 
\left(1- c_{j+}^\dag c_{j+} - c_{j-}^\dag c_{j-} 
\right) \; , \qquad
S_j^+ =  c_{j-}^\ds c_{j+}^\ds 
 \; , \qquad 
S_j^- = c_{j+}^\dag c_{j-}^\dag 
 \; , 
\end{eqnarray}
which satisfy the standard $SU(2)$ relations
\begin{eqnarray}
  \label{eq:SU2-relations}
 [ S_i^+, S_j^-] = \delta_{ij} \, 2 \, S_j^z \; , \qquad
 [ S_i^z, S_j^{\pm}] = \pm \delta_{ij} S_j^\pm \; , 
\end{eqnarray}
and in terms of which $H_U$ takes the form
\begin{eqnarray}
  \label{eq:HUspin-variables}
  H_U  & = &  \sum_j^U 2 \varepsilon_j (1/2- S_j^z) \, - \, 
g \sum_{ij} S_i^- S_j^+ \; .
\end{eqnarray}
Our choice (\ref{eq:pseudospin}) for the
  pseudospin variables differs from that used in many other
  publications\cite{CRS97,AFF,ADLO1} by the replacement 
\begin{equation}
\label{eq:otherconvention}
S_i^+ \leftrightarrow
  S_i^- \; , \qquad  S_i^z \rightarrow -S_i^z \; , 
\end{equation}
which preserves the $SU(2)$
relations (\ref{eq:SU2-relations}).  With our choice, the physical
  vacuum state $|0 \rangle$, containing no pairs, has
  the maximum possible $S_z$-eigenvalue and hence is a ``highest-weight''
  state. This is convenient for our present
  purpose, namely to establish contact with the ABA, because in the
  Bethe-Ansatz literature it is standard practice to use highest-weight
  states as reference states [see Eq.~(\ref{ket}) below].

Now, Richardson showed that the sought-after $P$-pair eigenstates
(unnormalized) and eigenenergies have the general form\cite{Sierra}
\begin{eqnarray}
  \label{eq:truebosoneigenstates-cc}
  | \Psi_P \rangle &=& \prod_{l=1}^P 
S^-(\mu_l) |0\rangle \, , \qquad 
\mbox{with} \quad 
  S^-(\mu_l) = \sum_i^U 
{S_i^- \over 2 \varepsilon_i - \mu_l} \; ,
\\
  \label{eq:truebosoneigenvalues}
\E_P &=& \sum_{l = 1}^P \mu_l \; .
\end{eqnarray} 
Here the $P$ parameters $\mu_l$ ($l  = 1, \dots , P$) are a
solution of a set of $P$ coupled algebraic equations,
which we shall call the ``Richardson equations'',
\begin{eqnarray}
  \label{eq:richardson-eigenvalues-cc}
  {1\over g }  - \sum_i^U 
{1  \over 2 \varepsilon_i - \mu_l}    
+ 
\sum_{\stackrel{l'=1}{l' \ne l}}^P 
{2 \over \mu_{l'} - \mu_l} = 0 \; ,
\qquad \mbox{for}\quad l  = 1, \dots, P \; .
\end{eqnarray}
These are to be solved (numerically,
see Appendix B of Ref.~\onlinecite{vDR}) subject to the restrictions $\mu_{l'
}\neq \mu_{l}$ if $l'  \neq l$.  A simple proof of this result may be found 
in Appendix B of Ref.~\onlinecite{vDR}; its strategy is to verify that $( H_U
- \E_P) |\Psi_P \rangle = 0$ by simply commuting $ H_U$ to the very
right past all of the $S_{l}^-$ operators 
in (\ref{eq:truebosoneigenstates-cc}).

Moreover, Cambiaggio, Rivas
and Saraceno \cite{CRS97} showed that 
the constants of the motion of $H_U$ have the form\cite{otherconvention} 
\begin{eqnarray}
  \label{eq:constants-of-motion-DBCS}
  {\bf H_i}= S_i^z + g 
\sum_{\stackrel{j=1}{j \ne i}}^U 
{ S_i^z S_j^z + \frac{1}{2}(S_i^+ S_j^- + S_i^- S_j^+) 
\over \varepsilon_i - \varepsilon_j} \; .
\end{eqnarray}
The operators ${\bf H_i}$, $i=1,2,\ldots ,N$ commute with 
each other as well as with the Hamiltonian (\ref{eq:HUspin-variables}).
In the limit $g\rightarrow \infty$, the operators 
\beq
\label{gaudinham}
{\bf H}_{\bf {i}}^{\rm Gaudin}=\lim_{g \to \infty} {\bf H_i}/g = 
\sum_{\stackrel{j=1}{j \ne i}}^U
{ S_i^z S_j^z + \frac{1}{2} (S_i^+ S_j^- + S_i^- S_j^+ )
\over \varepsilon_i - \varepsilon_j} \;
\eeq
coincide with the Hamiltonians of the Gaudin chain (see chapter 13 of 
Ref.~\onlinecite{gaudinbook}). The 
common eigenstates of the Gaudin Hamiltonians are given by the same 
Eqs.~(\ref{eq:truebosoneigenstates-cc}), but with the parameters $\mu_l$ 
satisfying the so-called Gaudin equations,
which are simply the $g\rightarrow \infty$ limiting case of Richardsons 
Eqs.~(\ref{eq:richardson-eigenvalues-cc}). The corresponding eigenvalues 
of the Hamiltonians ${\bf {H}}_{\bf i}^{\rm Gaudin}$ are given by
\beq
\label{gaudineigenvalue}  
{h}^{\rm Gaudin}_i=-\sum_{l=1}^P \frac{1}{2\varepsilon_i-\mu_l} + 
\frac{1}{2} \sum_{\stackrel{i'=1}{i' \ne i}}^U
\frac{2}{\varepsilon_i  - \varepsilon_{i'}} \; . 
\eeq 

\section{Algebraic Bethe Ansatz for the inhomogeneous 6-vertex model}
\label{ABA}

\subsection{Definition of Model}
\label{sec:define-model}

It is well known that the Gaudin model can be obtained by taking the
quasiclassical limit of the IXXX  model.\cite{gaudinbook}  The main
result of this paper will be to show that a similar construction can
be used to obtain the DBCS model from the TIXXX model, as well as
generalized DBCS models from the TIXXZ model.  To set the scene for
these developments, the next two sections give a pedagogical review of
the ABA as applied to IXXX and IXXZ models. Since both are special
cases of the so-called 6-vertex model, we begin by discussing the
latter in full generality. Bethe Ansatz experts may want to skip
directly to section~\ref{sec:Sklyanin}.

\begin{figure}
\begin{center}
{\includegraphics[clip,width=0.8\linewidth]{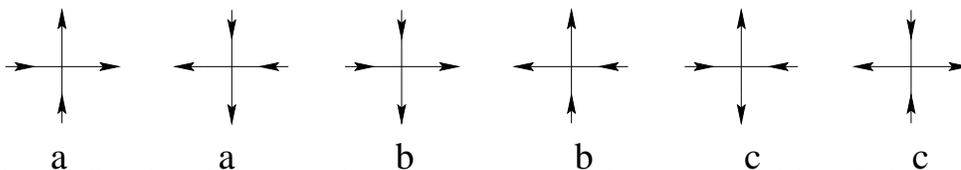}}
\caption{The six allowed configurations of arrows at a vertex of
the 6-vertex model, with their corresponding Boltzmann weigths $a$, $b$ and 
$c$. }
\label{fig:config}
\vspace{0.5cm}
\end{center}
\end{figure}

The 6-vertex model is a classical statistical mechanics model on a
two-dimensional regular quadratic lattice, whose dynamical variables are
arrows living along the horizontal and vertical edges
of the lattice, labelled by $m
= 1, \dots, M$ and $i = 1, \dots, N$, respectively.  At each vertex, only
those 6 configuration of arrows are ``allowed'', i.e.\ have nonzero local
Boltzman Weights (BW), for which the total flux into the vertex is zero
(see Fig.\ref{fig:config}).
Thus, every allowed configurations has exactly two incoming and two out-coming
arrows. Furthermore, we take the local BW's to be invariant with respect to
the simultaneous reflection of all four arrows.  This leaves only three
independent BW's per vertex, to be denoted by $a_{mi}$, $b_{mi}$ and $c_{mi}$,
where the subscripts give the location $(m,i)$ of the vertex (intersection of
row $m$ and column $i$).  Since the BW's are allowed to depend on the location
of the vertex, we are considering an ``inhomogeneous'' model.  As usual, the
total statistical weight of any given configuration is defined as the product
of the BW's of all vertices, and the partition function is defined as the sum
of these statistical weights over all qallowed configurations.

It is convenient to associate a two dimensional vector space ${\bf C}^2$ with
each row, say $U_m$ for row $m$, and another
with each column, say $V_i$ for column $i$,
in such a way that the basis vectors
\begin{mathletters}
\begin{eqnarray}
  \label{eq:basisvectors-row}
e^{(m)}_1 \; \equiv \;  {1 \choose 0} \; & \equiv \; \rightarrow \; , \qquad
 e^{(m)}_2  \;  &\equiv \; {0 \choose 1} \; \equiv \; \leftarrow \; , \\
\label{eq:basisvectors-column}
e^{(i)}_1 \; \equiv \; {1 \choose 0} \; & \equiv \; \uparrow \; , \qquad
e^{(i)}_2 \; & \equiv \; {0 \choose 1} \; \equiv \; \downarrow \; , 
\end{eqnarray}
\end{mathletters}
represent right- and left-pointing arrows along row $m$, 
or  upward and downward arrows
along column $i$, respectively.
Then the local BWs at vertex $(m,i)$ 
may be viewed as the matrix elements of a
linear operator ${\bf R_{mi}}$ that acts as follows 
on the tensor product of the $m$-th ``horizontal'' and $i$-th ``vertical''
 space, $U_m \otimes V_i$: 
\begin{mathletters}
\beq
\Rmi \, e^{(m)}_l  \otimes e^{(i)}_k 
= (\Rmi)^{\bar l}{}_l{}^{\bar k}{}_k \, 
e^{(m)}_{\bar l}  \otimes e^{(i)}_{\bar k} \; ,
\label{roperator} 
\eeq    
where the usual covention of summation over repeated indices 
$\bar l, \bar k \in \{1,2\}$ is implied. 
[As a rule, we shall put bars over all repeated indices,
and tilde's or nothing over non-repeated upper or lower indices,
respectively.] It follows that 
the action of $\Rmi$  on the coordinates $(\wmi)^{lk}$ of 
a general vector $\wmi \equiv 
(\wmi)^{\bar l \bar k}\,  e^{(m)}_{\bar l}  \otimes e^{(i)}_{\bar k}
 \in U_m \otimes V_i$ 
takes the form
\beq
(\Rmi \wmi)^{\tilde   l \tilde   k}
=   (\Rmi)^{\tilde   l}{}_{ \bar l}{}^{\tilde k}{}_{ \bar k} (\wmi)^{
  \bar l \bar k} \; . 
\eeq  
\end{mathletters}  
The only nonzero matrix  
elements of the operator $\Rmi$ are
\beq
& & (\Rmi)^1{}_1{}^1{}_1 \; = \; (\Rmi)^2{}_2{}^2{}_2 \; = \; a_{mi} \; , 
\nonumber \\
& & (\Rmi)^1{}_1{}^2{}_2 \; = \; (\Rmi)^2{}_2{}^1{}_1 \; = \; b_{mi} \; , 
\label{rmat}
\\
& & 
(\Rmi)^1{}_2{}^2{}_1\; = \;  (\Rmi)^2{}_1{}^1{}_2  \; = \; c_{mi} \; .
\nonumber 
\eeq
A convenient  matrix representation for ${\bf R_{mi}}$ is 
\begin{mathletters}
\beq
\label{eq:R-indices} ({\bf R_{mi}})^{\tilde   l}{}_{l}{}^{\tilde k}{}_{k}
= \left( \begin{array}{cc} 
(\bm{\alpha}_{\bf mi})^{\tilde k} {}_k  & (\bm{\beta}_{\bf mi})^{\tilde k} {}_k 
\\
(\bm{\gamma}_{\bf mi})^{\tilde k} {}_k   & (\bm{\delta}_{\bf mi})^{\tilde k} {}_k 
\end{array} \right)^{\tilde l}_{\; \; l}  \; , 
\eeq
where $\bm{\alpha}_{\bf mi}$, $\bm{\beta}_{\bf mi}$,
 $\bm{\beta}_{\bf mi}$, $\bm{\gamma}_{\bf mi}$
are operators acting on the two dimensional vertical space $V_i$:
\beq
\bm{\alpha}_{\bf mi} = {a_{mi} \quad  0  \; \choose \; 0 \quad b_{mi} } ,
\quad
\bm{\beta}_{\bf mi} = { \; 0  \quad  0  \choose c_{mi} \quad 0 \; } ,
\quad
\bm{\gamma}_{\bf mi} = {\; 0  \quad  c_{mi}  \; \choose  0 \; \quad 0 \; } ,
\quad
\bm{\delta}_{\bf mi}= {b_{mi} \quad  0  \; \choose \; 0 \quad a_{mi} } . 
\eeq
\end{mathletters}
Even more explicitly, $\Rmi$ can be expressed 
as follows in terms of the unit operator ${\bf I}$ and the 
Pauli $\bm{\sigma}$-matrices $\bm{\sigma}^z$, $\bm{\sigma}^{\pm}=
( \bm{\sigma}^x \pm i\bm{\sigma}^y)/2$,  
\beq 
\Rmi =\frac{a_{mi}+b_{mi}}{2}\left( {\bf I_m} \otimes {\bf I_1} \right)
+ \frac{a_{mi}- b_{mi}}{2}
\left( \bm{\sigma}_{\bf m}^z \otimes \bm{\sigma}_{\bf i}^z \right)
+c_{mi} \left(\bm{\sigma}_{\bf m}^+ \otimes \bm{\sigma}_{\bf i}^- + 
\bm{\sigma}_{\bf m}^- \otimes \bm{\sigma}_{\bf i}^+\right) \; ,
\label{rmatpauli}  
\eeq
where  the lower indices of the operators indicate the space 
($U_m$ or $V_i$) on which  they act. 

\subsection{Monodromy Matrix}
\label{sec:monodromy}

\begin{figure}
\begin{center}
  {\includegraphics[clip,width=0.70\linewidth]{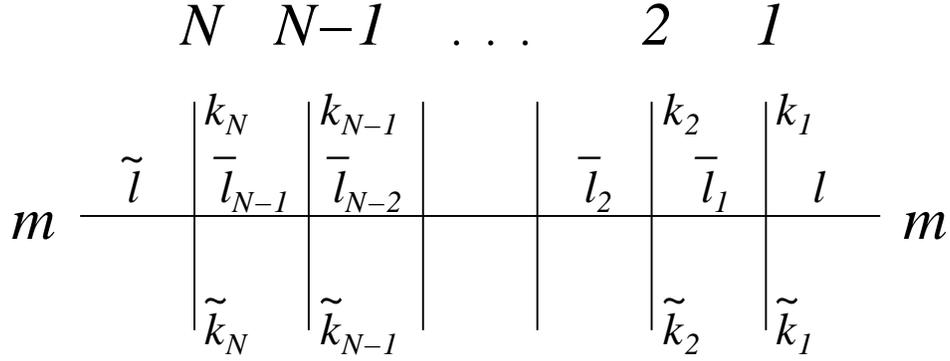}}
\vspace{1cm}
\caption{Construction of the Monodromy matrix:
  the matrix element $({\cal T}_m)^{\tilde l,}_{\; \; l,} {}^{\tilde k_N \dots
    \tilde k_1}_{\; \; k_N \dots k_1}$ is equal to the total Boltzmann weight of
  the $m$-th row, for a fixed configuration of external arrows [specified by
  the indices of ${\cal T}_m$], obtained by
  summing over all allowed configurations of arrows on internal lines
(whose indices carry bars here).  }
\label{fig:mmat}
\vspace{0.5cm}
\end{center}
\end{figure}

One of the most important objects in the ABA
method is the Monodromy matrix ${\cal T}_m$. It is
defined to be the operator
\begin{mathletters}
\beq
{\cal T}_{m} = {\bf R_{mN}\; R_{mN-1} \ldots R_{m1}} \; ,
\label{mmat} 
\eeq 
which acts on the space $U_m \otimes V_1 \ldots
\otimes V_N$, with each factor $\Rmi$ acting nontrivially only on the
``horizontal'' space $U_m$ and the ``vertical'' space $V_i$.
To illustrate this action explicitly, we note that 
the matrix elements of ${\cal T}_m$ are constructed
as follows from those of ${\bf R_{mi}}$:
 \begin{eqnarray}
   \label{eq:monodromy-explicit}
   ({\cal T}_m)^{\tilde l,}_{\; \; l,}
{}^{\tilde k_N \dots \tilde k_1}_{\; \; k_N \dots k_1}
= ({\bf R_{mN}})^{\tilde l}_{\; \; \bar l_{N-1}}
{}^{\tilde k_N}_{\; \; k_N} 
({\bf R_{mN-1}})^{\bar l_{N-1}}_{\; \; \bar l_{N-2}}
{}^{\tilde k_{N-1}}_{\; \; k_{N-1}} \dots
({\bf R_{m1}})^{ \bar l_1}_{\; \; l}
{}^{\tilde k_{1}}_{\; \; k_{1}} \; .
 \end{eqnarray}
\end{mathletters}
This equation  has a simple physical interpretation: 
each such matrix element of ${\cal T}_m$
 gives the total Boltzmann weight of the $m$-th row, depicted in
 Fig.\ref{fig:mmat}, for a fixed configuration of external arrows
 (specified by the indices of ${\cal T}_m$ on the left-hand side of
 \Eq{eq:monodromy-explicit}), obtained by summing over all allowed
 configurations of arrows on internal horizontal edges (the $\bar l_i$
 sums, for $i = 2, \dots, N$, on the right-hand side of
 \Eq{eq:monodromy-explicit}). Likewise, using this one row
construction as building block, the partition function of an $M$-row
lattice can be expressed via the matrix elements of a suitable product
of the $M$ Monodromy matrices, as will be seen below.

Because of the different roles played by the horizontal space $U_m$ 
(usually reffered as the auxiliary space) and the tensor product of 
remaining vertical  spaces $V_1 \cdots V_N$ (the so called quantum 
space), it is convenient to arrange the matrix elements of ${\cal T}_m$ 
that correspond to the horizontal space $U_m$, i.e.\
$({\cal T}_m)^{\tilde l}_{\; l}$ in the notation of
\Eq{eq:monodromy-explicit}, into a $2 \times 2$
matrix:
\beq
{\cal T}_m \equiv \left( \begin{array}{c c} A_m & B_m \\ C_m & D_m 
\end{array}\right) \; .
\label{mmat2x2} 
\eeq
Its entries $A$, $B$, $C$ and $D$ are, of
course, operators acting on the quantum space, which 
implicitly carry the $k$-indices that are displayed in
\Eq{eq:monodromy-explicit} (for brevity, we suppressed these
above).  Each of these four matrix elements corresponds to one of four
possible kinds of rows in Fig. 2, depending on how the arrows on the
first and last (i.e.\ external) horizontal edges are fixed:
\begin{eqnarray}
  \label{eq:edges-for-mmat}
  A:  (\rightarrow, \rightarrow) , \qquad
  B:  (\rightarrow, \leftarrow) , \qquad
  C:  (\leftarrow, \rightarrow) , \qquad
  D:  (\leftarrow, \leftarrow) .
\end{eqnarray}
Note that in this matrix representation, the product of ${\bf R_{mi}}$
matrices on the r.h.s of Eq.~(\ref{mmat}) may be viewed as
conventional multiplication of $2\times 2$ matrices [of the form
(\ref{eq:R-indices})], whose entries are,
however, operators on the quantum space (and hence carry suppressed
$k$-indices).

\subsection{Transfer Matrix}
\label{sec:transfer-matrix}

In order to investigate our model in the case of \emph{periodic boundary
  conditions} in the horizontal direction, it is natural to consider the
operator $\Tm$ (called transfer matrix), defined as the trace of ${\cal T}_m$
in the horizontal space $U_m$: 
\beq \Tm \; \equiv \; {\rm Tr}_m \{ {\cal T}_m \} \; \equiv \; \sum_{l=1,2}
({\cal T}_m)^l_{\; l} \; = \;  
A_m + D_m \; .
\label{tmat}
\eeq Its matrix elements $(\Tm)^{\tilde k_N \dots \tilde k_1}_{\; \;
  k_N \dots k_1}$ give the total Boltzmann weight of the $m$-th row
depicted in Fig. 2 for a fixed configuration of arrows on the vertical
edges [specified by the indices of $\Tm$], obtained by summing over
all allowed configurations of arrows on horizontal edges, with the
boundary condition that the arrows of the first and last horizontal
edges are equal. 

It follows that the full partition function for a lattice of $M$ rows
and $N$ columns can readily be constructed by a suitable
product of $M$ transfer matrices: for the case of double periodic boundary
conditions, it is equal to 
\beq 
Z_{M,N} & \equiv & ({\bf
  T_M})^{\bar k^{M}_N \dots \bar k^{M}_1}_{\; \; \bar k^{M-1}_N \dots \bar k^{M-1}_1}
({\bf T_{M-1}})^{\bar k^{M-1}_N \dots \bar k^{M-1}_1}_{\; \; \bar k^{M-2}_N \dots
  \bar k^{M-2}_1} \dots ({\bf T_1})^{\bar k^{1}_N \dots \bar k^{1}_1}_{\; \; \bar k^{M}_N
  \dots \bar k^{M}_1}
\\
& \equiv & {\rm Tr} \{ {\bf T_M} {\bf T_{M-1}} \dots {\bf T_1} \} \; ,
\label{partfunc}
\eeq 
i.e.\ the trace is over the entire quantum space $V_1\otimes \ldots V_N$.

The distinguishing feature of \emph{integrable} models
is that the transfer matrices for different rows commute, 
${\bf T_m} {\bf T_{m'}} = {\bf T_{m'}} {\bf T_m}$.
In this case, all transfer matrices have common eigenstates,
with eigenvalues $\Lambda_m^{(\alpha)}$, say, so that the partition
function takes the form
\begin{eqnarray}
  \label{eq:partition-function-product}
  Z_{M,N} = \sum_\alpha \prod_{m=1}^M \Lambda_m^{(\alpha)} \; .
\end{eqnarray}
Thus, the calculation of the partition function reduces to the problem
of finding the eigenvalues of the transfer matrices.

\subsection{Yang-Baxter Relations}

It turns out (and will be shown below) that the transfer matrices
commute if the local 
BW's $a_{mi}$, $b_{mi}$, $c_{mi}$ are parametrized as follows:
\begin{mathletters}
\label{eq:phi-Ansatz}
\beq 
&& a_{mi} =1 \; ; \qquad b_{mi} = b(\l_m ,\xi_i) \; ; 
\qquad c_{mi} = c (\l_m ,\xi_i) \; , 
\label{abc}
\\ 
\label{eq:phi}
&& b(\l , \xi) = \frac{\phi(\l -\xi )} {\phi(\l -\xi +2\eta)} \; ,
\qquad c(\l, \xi) = \frac{\phi(2\eta )}{\phi(\l -\xi +2\eta)}, 
\eeq 
\end{mathletters}
where form of the function $\phi(x)$ can be either $\phi(x) =x$ 
or $\phi(x)=\sinh x$.
The parameter $\l_m$
(called ``spectral parameter'') is associated with the $m$-th horizontal line,
and $\xi_i$ (the inhomogeneity parameter) with the $i$-th 
vertical line.  Note that the
dependence of the local BW's $a_{mi}$, $b_{mi}$, $c_{mi}$ on their indices
thus enters only via their spectral parameters, which is why we could
introduce functions $b(\lambda,\xi)$ and $c(\lambda, \xi)$ that don't carry
the indices $(m,i)$ any longer.
Note also that the ratio $c/b$ is antisymmetric under 
an interchange of its arguments:
\beq
\label{eq:c/b-antisymmetry}
 {c(\lambda, \xi) \over b(\lambda, \xi) } = - 
{c(\xi, \lambda) \over b(\xi, \lambda) } \; ,
\eeq
a property that will be useful later.

Usually the rational case $\phi(x)=x$ is referred to as the XXX model and the
trigonometric case $\phi(x)=\sinh x$ as the XXZ model, since the Hamiltonians
of the XXX and XXZ Heisenberg magnetic chains can be derived from the
corresponding (homogeneous) transfer matrices, by taking a logarithmic
derivative with respect to the spectral parameter at some specific point (see
e.g. chapter 10.14 of Ref.~\onlinecite{baxterbook}).

\begin{figure}
\begin{center}
{\includegraphics[clip,width=0.7\linewidth]{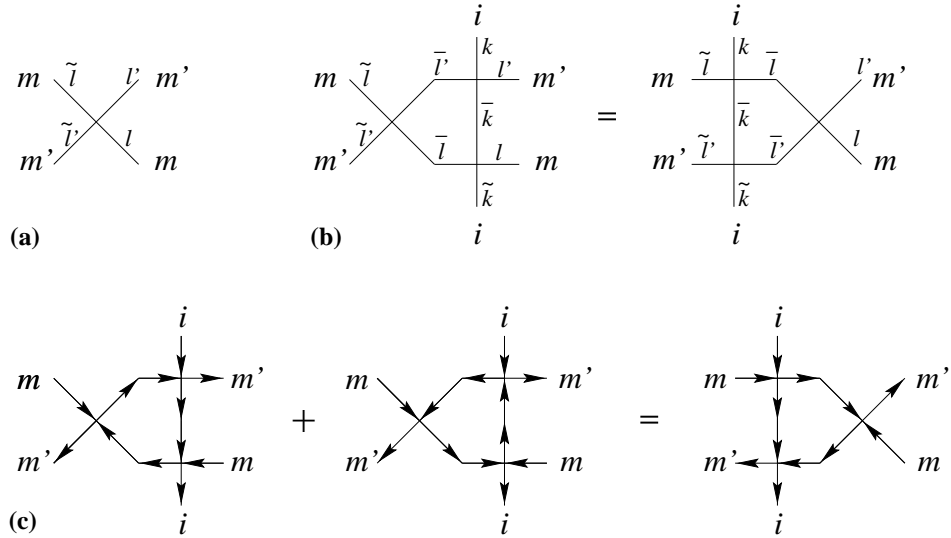}}
\caption{Graphical depiction of the Yang-Baxter equations:
(a): Schematic depiction of the action of 
$({\bf \tilde R_{mm'}})^{\tilde l}_{\; \;  l}{}^{\tilde l'}_{\; \; l'}$,
which  interchanges the order of the rows $m'$ and $m$.
(b): General graph for the Yang-Baxter equation (\ref{ybe-explicit}).
(c): Specific  graph for a particular configuration
of external arrows, representing the specific
Yang-Baxter equation (\ref{eq:YB-check-explicit}). 
Summing over all possible configurations of internal arrows
consistent with the given choice of external arrows
turns out to give two graphs on the left-hand side,
but only one on the right-hand side.}
\label{fig:ybe}
\vspace{0.5cm}
\end{center}
\end{figure}

For the choice of BW's of  \Eqs{eq:phi-Ansatz},
the $\Rmi$-matrices have the following very important
property, which ultimately leads to the solution of the problem:
they satisfies the Yang-Baxter (YB) equation 
\begin{mathletters}
\beq
{\bf \tilde R_{mm'}}(\l_{m}, \l_{m'}) \, {\bf R_{mi}} (\l_{m}, \xi_i) \, 
{\bf R_{m'i}} (\l_{m'}, \xi_i) = 
{\bf R_{m'i}} (\l_{m'}, \xi_i) \, {\bf R_{mi}} (\l_{m}, \xi_i) \, 
{\bf \tilde R_{mm'}} (\l_{m}, \l_{m'}) \; ,
\label{ybe}
\eeq where the operator products on both sides act on the space $U_m
\otimes U_{m'} \otimes V_i$, and the arguments in brackets indicate
explicitly on which parameters the corresponding operators depend. As
before, the ${\bf R_{mi}}$ operators act on one horizontal and one
vertical space, $U_m \otimes V_i$; their nonzero matrix elements are
given in Eq.  (\ref{rmat}), with the parameters $a_{mi}$, $b_{mi}$ and
$c_{mi}$ as defined in \Eq{abc}, with arguments $\l_m$ and $\xi_i$.
In contrast, the operator ${\bf \tilde R_{mm'}}$ acts on \emph{two
  horizontal} spaces, $U_m$ and $U_{m'}$; apart from this replacement
of vertical space $V_i$ by the horizontal space $U_{m'}$, however, the
structure of ${\bf \tilde R_{mm'}}$ 
is exactly the same as that of $\Rmi$: the nonzero matrix
elements of ${\bf \tilde R_{mm'}}$ are likewise given by Eq.
(\ref{rmat}), where now the parameters $a_{mm'}$, $b_{mm'}$ and
$c_{mm'}$ have arguments $\l_{m}$ and $\l_{m'}$ (i.e.\ two $\lambda$'s
instead of $\lambda$ and $\xi$): \beq && a_{mm'} =1 \; ; \qquad
b_{mm'} = b(\l_{m} , \l_{m'}) \; ; \qquad c_{mm'} = c (\l_{m} ,
\l_{m'}) \; .
\label{abc-xi-lambda} \nonumber
\eeq
To be explicit, the Yang-Baxter equation implies the 
following relations between matrix elements of the transfer
matrices:
\beq
({\bf \tilde R_{mm'}})^{\tilde l}_{\; \; \bar l}{}^{\tilde l'}_{\; \; \bar l'}
 \, ({\bf R_{mi}})^{\bar l}_{\; \; l}{}^{\tilde k}_{\; \; \bar k} \,
({\bf R_{m'i}})^{\bar l'}_{\; \; l'}{}^{\bar k}_{\; \; k}
\;  = \;
({\bf R_{m'i}})^{\tilde l'}_{\; \; \bar l'}{}^{\tilde k}_{\; \; \bar k}
 \, ({\bf R_{mi}})^{\tilde l}_{\; \; \bar l}{}^{\bar k}_{\; \; k}
  \, ({\bf \tilde R_{mm'}})^{\bar l}_{\; \; l}{}^{\bar l'}_{\; \; l'} \; .
\label{ybe-explicit}
\eeq 
\end{mathletters}
(As usual, we used bars for repeated indices, which are summed over).
Graphically, this equation can be represented as the equality of the BW's of
the configurations depicted in Fig.~\ref{fig:ybe}(b), 
where the left and right diagrams have
the \emph{same} configuration of arrows on all external edges, and sums over
all possible configurations of internal indices are implied.  The verification
of the YB equation (\ref{ybe}) is straightforward (though rather tedious) and
reduces to some simple rational or trigonometric identities. (Actually, the
solvability of these Yang-Baxter equations dictated the choice of
parametrization of $a$, $b$ and $c$ made in \Eq{eq:phi}.)  As an illustration,
the graph in Fig.~3(b) corresponds to the following explicit realization of
\Eq{ybe-explicit}:
\begin{mathletters}
\label{eq:YB-check-explicit}
\begin{eqnarray}
\nonumber
& & ({\bf \tilde R_{mm'}})^{1}_{\; \; 2}{}^{2}_{\; \; 1}
 \, ({\bf R_{mi}})^{2}_{\; \; 2}{}^{2}_{\; \; 2} \,
({\bf R_{m'i}})^{1}_{\; \; 1}{}^{2}_{\; \; 2} \; + \;
({\bf \tilde R_{mm'}})^{1}_{\; \; 1}{}^{2}_{\; \; 2}
 \, ({\bf R_{mi}})^{1}_{\; \; 2}{}^{2}_{\; \; 1} \,
({\bf R_{m'i}})^{2}_{\; \; 1}{}^{1}_{\; \; 2}
\\ && \qquad \qquad \qquad \;  = \;
({\bf R_{m'i}})^{2}_{\; \; 2}{}^{2}_{\; \; 2}
 \, ({\bf R_{mi}})^{1}_{\; \; 1}{}^{2}_{\; \; 2}
  \, ({\bf \tilde R_{mm'}})^{1}_{\; \; 2}{}^{2}_{\; \; 1} \; ,
\end{eqnarray}
or, using \Eqs{rmat}, 
\begin{eqnarray}
\label{eq:YB-check}
c_{mm'}a_{mi}b_{m'i}+
b_{mm'}c_{mi}c_{m'i}= 
a_{m'i}b_{mi}c_{mm'} \; , 
\end{eqnarray}
\end{mathletters}
which can be verified to hold if $a_{mi}$, $b_{mi}$
and $c_{mi}$ have the form specified in \Eqs{eq:phi-Ansatz}.

Now consider two Monodromy matrices ${\cal T}_m$
and ${\cal T}_{m'}$ with identical sets of inhomogeneity 
parameters $\xi_1, \dots \xi_N$,
\begin{eqnarray}
  \label{eq:monodromy-lambda}
 {\cal T} (\l_m) \equiv {\cal T}_m  (\l_m ; \xi_1 , \ldots \xi_N) 
 \; , 
\qquad
{\cal T} (\l_{m'}) \equiv
{\cal T}_{m'} (\l_{m'}; \xi_1 , \ldots \xi_N) \; , 
\end{eqnarray}
so that it suffices to display 
only the functional dependence on the (arbitrary) spectral parameters
$\lambda_m$ or
$\lambda_{m'}$. For this case,
we shall  write the components of ${\cal T}(\l_m)$ as
$A(\l_m)$, $B(\l_m)$, $C(\l_m)$ and $D(\l_m)$, and the transfer matrix as
${\bf T} (\l_m)$, while 
using ${\bf \tilde R_{mm'}}$ as shorthand for ${\bf \tilde R_{mm'}} (\l_{m},
\l_{m'})$.
A direct consequence of the YB equation is that 
two such Monodromy matrices satisfy
the following exchange relations,
\begin{mathletters}
\beqa {\bf \tilde R_{mm'}} {\cal T} (\l_{m} ) {\cal T} (\l_{m'}) \; = \; {\cal
  T} (\l_{m'}) {\cal T} (\l_{m} ) {\bf \tilde R_{mm'}} \; ,
\label{mmexr}
\eeqa 
or, in terms of matrix elements, 
\beqa 
\label{mmexr-specific}
({\bf \tilde R_{mm'}})^{\tilde l}_{\; \bar l} {}^{\tilde
  l'}_{\; \bar l'} {\cal T}^{\bar l,}_{\; \; l,}{}^{\tilde k_N \dots \tilde
  k_1}_{ \; \; \bar k_N \dots \bar k_1} (\l_{m}) {\cal T}^{\bar l',}_{\; \;
  l',} {}^{\bar k_N \dots \bar k_1}_{\; \; k_N \dots k_1} (\lambda_{m'}) \; =
\; {\cal T}^{\tilde l',}_{\; \; \bar l',}{}^{\tilde k_N \dots \tilde k_1}_{ \;
  \; \bar k_N \dots \bar k_1} (\l_{m'}) {\cal T}^{\tilde l,}_{\; \; \bar
  l,}{}^{ \bar k_N \dots \bar k_1}_{\; \; k_N \dots k_1} (\l_{m}) ({\bf \tilde
  R_{mm'}})^{\bar l}_{\; l} {}^{\bar l'}_{\; l'} \; . 
\eeqa 
\end{mathletters}
Fig. 4  is a
graphical representation of this relation. To prove it, one successively
``pulls'' the crossing of the two horizontal lines across from the right-most
edge of the quantum space to the left-most edge, using the graphical 
representation Fig. 3 of the YB equation (\ref{ybe}).
\begin{figure}
\begin{center}
{\includegraphics[clip,width=0.98\linewidth]{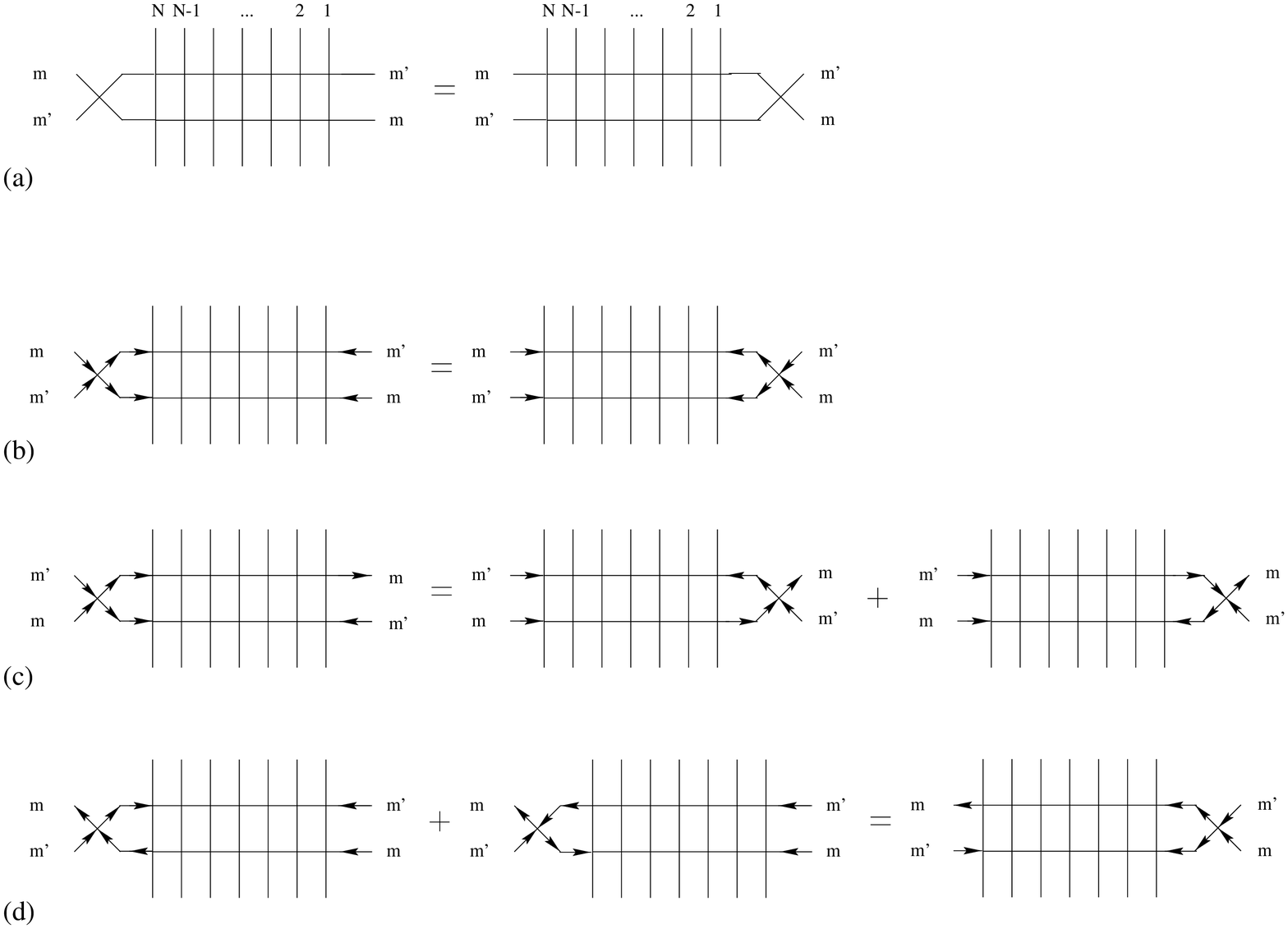}}
\caption{}
\label{fig:4} Illustration of the exchange relations
(\ref{mmexr})
for the Monodromy matrix. (a) The general relation (\ref{mmexr}).
(b), (c), (d): Three specific choices of external
arrows, leading to the three Eqs. (\ref{bbcommrel}),
(\ref{abcommrel}) and (\ref{dbcommrel}), respectively.
\vspace{0.5cm}
\end{center}
\end{figure}
Rewriting the Eq.~(\ref{mmexr}) 
in the form 
\beq
{\bf \tilde R_{mm'}} {\cal T} (\l_{m} )
{\cal T} (\l_{m'} )
{\bf \tilde R^{-1}_{mm'}} \; 
\; = \; {\cal T} (\l_{m'})  {\cal T} (\l_{m}) \;, 
\eeq
taking traces over the spaces $U_{m}$, $U_{m'}$ and using 
the cyclic property of the trace operation,  one immediately 
concludes that the corresponding transfer matrices commute:
\beq
{\bf T} (\l_{m}) {\bf T} (\l_{m'}) = 
{\bf T} (\l_{m'}) {\bf T} (\l_{m}) \; . 
\label{tmcomm}
\eeq
It is the existence of a one-parameter family of commuting transfer 
matices that makes the exact calculation of their eigenvalues and 
construction of their common eigenstates possible.

\section{Eigenstates and eigenvalues of the transfer matrix}
\label{sec:eigenstates}

Eq.(\ref{mmexr}) represents in a compact form 16 
commutation relations among the matrix entries
$A(\l_m)$, $B(\l_m)$, $C(\l_m)$ and $D(\l_m)$ of
the Monodromy matrix ${\cal T} (\lambda_m)$ 
(see Eq.~(\ref{mmat2x2})). Below we wright down 
three of them, 
which are essential for solving our eigenvalue 
eigenstate problem (the full set of relations can be found, e.g.\, in 
chapter VII of Ref.~\onlinecite{izerginkorepin}): 
\begin{mathletters}
\label{commrel}
\beq
\left[ B(\l_m), B(\l_{m'}) \right] &=&0 
\label{bbcommrel}
\\ 
A(\l_m) B(\l_{m'}) &=&\frac{1}{b(\l_{m'},\l_m)} 
B(\l_{m'}) A(\l_m)+ \frac{c(\l_m, \l_{m'})}
{b(\l_m, \l_{m'})} B(\l_m) A(\l_{m'}),
\label{abcommrel}
\\
D(\l_m) B(\l_{m'}) &=&\frac{1}{b(\l_m,\l_{m'})} 
B(\l_{m'}) D(\l_m)+ \frac{c(\l_{m'}, \l_m)}
{b(\l_{m'}, \l_m)} B(\l_m) D(\l_{m'}) \; . 
\label{dbcommrel}
\eeqa
\end{mathletters} 
These three equations correspond to  the graphical 
equations of Fig.~4(b), 4(c) and 4(d) (in that order); for
example, Fig.~4(d) represents the following specific
realization of \Eq{mmexr-specific} (whose $k$-indices
we here suppress):
\beqa 
&&
({\bf \tilde R_{mm'}})^{2}_{\; 2} {}^{1}_{\; 1} 
{\cal T}^{2}_{\; \; 2}
 (\l_{m}) {\cal T}^{1}_{\; \;   2,}  (\lambda_{m'}) \; 
+ \;
({\bf \tilde R_{mm'}})^{2}_{\; 1} {}^{1}_{\; 2} 
{\cal T}^{1}_{\; \; 2}
 (\l_{m}) {\cal T}^{2}_{\; \;   2}  (\lambda_{m'}) 
\\
&&\qquad \qquad \qquad \qquad \; =
\; {\cal T}^{1}_{\; \; 2} (\l_{m'}) 
{\cal T}^{2}_{\; \; 2} (\l_{m}) ({\bf \tilde
  R_{mm'}})^{2}_{\; 2} {}^{2}_{\; 2} \; . \eeqa
Using \Eqs{rmat} and the antisymmetry property
(\ref{eq:c/b-antisymmetry}), this readily reduces to 
\Eq{dbcommrel}.

Now we are ready to construct some eigenstates of the model.
As reference state we use the following so-called vacuum state
or highest-weight state,
\beq
\ket \equiv e^{(1)}_1\otimes \ldots \otimes e^{(N)}_1 \in V_1 \otimes 
\ldots \otimes V_N \; , 
\label{ket}
\eeq
which can be visualized as a row of vertical edges,
each of which carries an upward-pointing arrow.
It is easy to verify that 
\begin{mathletters}
\label{rstate}
\beq
C(\l_m) \ket &=&0,  
\label{crefstate}
\\
A(\l_m) \ket &=& \ket ,
\label{arefstate}  
\\
D(\l_m) \ket &=& \prod^N_{i=1} b(\l_m , \xi_i)\ket .
\label{drstate}
\eeq
\end{mathletters}    
To check, e.g., Eq.~(\ref{drstate}), recall
that the operator $D$ has the graphical representation shown in 
Fig. 2, with left-pointing arrows put on the first and last 
horizontal edges. The action on $\ket$ implies arranging upward-pointing 
arrows on all the vertical edges above the horizontal line, but then 
the only allowed arrangement of arrows on the edges below the 
horizontal line is again a sequence of exclulsively upward-pointing arrows.
By Eq.~(\ref{eq:monodromy-explicit}), this implies 
that we must have $N$ successive $b$-type vertices, as 
specified in  Eq.~(\ref{drstate}).

To obtain more general states, one can act on the reference state 
$\ket$ by an arbitrary number $P$ (with $1 \le P \le M$)
of operators $B$: 
\beq
|\mu_1 , \ldots , \mu^\pdag_P \rangle 
\equiv B(\mu_1) \ldots B(\mu^\pdag_P) \ket .
\label{betv}
\eeq
Note that the parameters $\mu_l$, $l = 1 , \dots, P$,
in the arguments of the $B$'s are arbitrary now, 
and unrelated to the  spectral parameters introduced earlier,
for which $\lambda_m$ is associated with row $m$.
We shall now show  that when these $\mu_l$ parameters are solutions of 
a particular system of equations (the famous Bethe-Ansatz equations, 
see Eq.~(\ref{baeq}) below), the vector (\ref{betv}) becomes an 
eigenstate of the transfer matrix ${\bf T}(\l)$ 
(we shall henceforth usually
drop the index $m$ on $\l_m$, since only the functional
dependence is important). 
To this end, let us analyze in
detail the action by the operator $D(\l)$ on the Bethe vector 
(\ref{betv}) (subsequently, the action of $A(\l)$, which is analogous, will
be outlined somewhat more briefly). Our strategy 
is simple: using the exchange relations (\ref{dbcommrel}), we move the 
operator $D$ to the right past all $B$'s until it appears 
next to the vacuum state 
$|0\rangle $, on which it acts according to (\ref{drstate}). The result can 
be represented as:
\begin{mathletters}
\label{daction-both}
\beqa
D(\l)|\mu_1, \ldots ,\mu^\pdag_P \rangle & = & \prod_{l=1}^P {1 \over 
b(\l , \mu_l) }
\prod_{i=1}^N b(\l,\xi_i)|\mu_1 , \ldots , \mu^\pdag_P \rangle 
\nonumber \\
& & + \sum_{l=1}^P f^D_l |\mu_1 , \ldots , \mu_{l-1},\l , \mu_{l+1} , 
\ldots   \mu^\pdag_P \rangle \; , 
\label{daction}
\\
f^D_l & =& \frac{c(\mu_l,\l)}{b(\mu_l,\l)} 
\prod_{\stackrel{l'=1}{l' \ne l}}^P
\frac{1}{b(\mu_l,\mu_{l'})}
\prod_{i=1}^Nb(\mu_l,\xi_i).
\label{dunwanted}
 \eeqa 
\end{mathletters}
The first (so-called ``wanted'') term on the r.h.s.\ of
Eq.~(\ref{daction}) arises from the case for which one picks
up,  at each of the series of commutations,
the first term of 
the exchange rule (\ref{dbcommrel}), which in our case is equal to   
\beq
\frac{1}{b(\l,\mu_l)} B(\mu_l) D(\l). 
\eeq
Below we will refer to this term in exchange relation
as a ``regular''term.
All other terms (the so called unwanted terms), which have contributions 
from the second term of (\ref{dbcommrel}), are combined in the second line 
of  Eq.~(\ref{daction}). The form of the coefficient $f^D_l$ occuring
in this term is given in Eq.~(\ref{dunwanted}) and 
can be derived as follows:
the l.h.s.\ of the Eq.~(\ref{daction}) is symmetric with respect to 
the permutations of the parameters $\mu_1, \ldots ,\mu^\pdag_P$,
due to the commutativity (\ref{bbcommrel}) of $B$'s. Since 
the first term on the r.h.s.\ of Eq.~(\ref{daction}) is symmetric 
as well, the second should be too. This means that if one succeedes 
to determine a single coefficient $f^D_l$ for some fixed $l$, then 
all other $f$'s can be straightforwardly found using the symmetry. Let us 
consider the case $l=1$. It is not difficult to see that the only 
possibility to obtain a term that is proportional to 
$|\l, \mu_2, \ldots ,\mu^\pdag_P \rangle $ and does not contain the
operator $B(\mu_1)$, is to choose the ``wrong'' term of (\ref{dbcommrel}),
\beq
\frac{c(\mu_1, \l)}{b(\mu_1, \l)} B(\l) D(\mu_1)  \; , 
\eeq
at the very first step when commuting $D(\l)$ with $B(\mu_1)$, and 
then everywhere else to choose ``regular'' ones. Thus, for $f^D_1$ 
we obtain:
\beq
f^D_1 =\frac{c(\mu_1,\l)}{b(\mu_1,\l)} \prod_{l=2}^P
\frac{1}{b(\mu_1,\mu_l)}
\prod_{i=1}^Nb(\mu_1,\xi_i).
\eeq
The above-mentioned symmetry under the permutations 
of $\mu_l$'s then immediately  implies that in general $f^D_l$
must have the form given in Eq.~(\ref{dunwanted}). 

A similar consideration of the action by the operator $A$ on the 
Bethe vector (\ref{betv}) gives:
\begin{mathletters}
\label{aaction-both}
\beq
A(\l)|\mu_1, \ldots ,\mu^\pdag_P \rangle & = & 
\prod_{l=1}^P
\frac{1}{b(\mu_l,\l)}
|\mu_1 , \ldots , \mu^\pdag_P \rangle 
+ \sum_{l=1}^p f^A_l |\mu_1 , \ldots , \mu_{l-1},\l , \mu_{l+1} , 
\ldots   \mu^\pdag_P \rangle ,
\label{aaction}
\\
f^A_l & = & \frac{c(\l,\mu_l)}{b(\l,\mu_l)} 
\prod_{\stackrel{l'=1}{l' \ne l}}^P
\frac{1}{b(\mu_{l'},\mu_l)} .
\label{aunwanted}
\eeq
\end{mathletters}
Combining Eqs. (\ref{daction-both}) and
(\ref{aaction-both}), we see that  
the state (\ref{betv}) is an eigenstate of the 
transfer matrix ${\bf T}(\l)=A(\l)+D(\l)$ 
\begin{mathletters}
\beq
{\bf T}(\l) |\mu_1 , \ldots , \mu^\pdag_P \rangle = 
t(\l ;\mu_1 , \ldots , \mu^\pdag_P ) |\mu_1 , \ldots , \mu^\pdag_P \rangle 
\eeq
with the eigenvalue
\beq
t(\l ;\mu_1 , \ldots , \mu^\pdag_P )=
\prod^P_{l=1} {1 \over b (\mu_l ,\l)} + 
\prod^P_{l=1} {1 \over b (\l, \mu_l )} \prod^N_{i=1} b(\l , \xi_i) \; , 
\label{eigenv} 
\eeq
\end{mathletters}
provided that 
$f^A_l+f^D_l=0$
 for every $l=1,2, \ldots P$. Using
the antisymmetry property of the ratio $c/b$,
\Eq{eq:c/b-antisymmetry},
it is easy to see that this condition 
is satisfied provided that the  $P$ 
parameters $\mu_1, \ldots ,\mu_P$ satisfy 
the following system of $P$ equations, 
\beq
\prod_{\stackrel{l'=1}{l' \ne l}}^P 
\frac{b(\mu_l, \mu_{l'})}{b(\mu_{l'}, \mu_l)} = 
\prod_{i=1}^N b(\mu_l, \xi_i) \; , 
\label{baeq}
\eeq
which are known as the Bethe equations. Every solution of the system 
of Bethe equations defines an eigenstate and corresponding 
eigenvalue of the transfermatrix ${\bf T}(\l)$ via (\ref{betv}) 
and (\ref{eigenv}).  

\section{Sklyanin's K-matrix}
\label{sec:Sklyanin}

\begin{figure}[t]
\begin{center}
{\includegraphics[clip,width=0.8\linewidth]{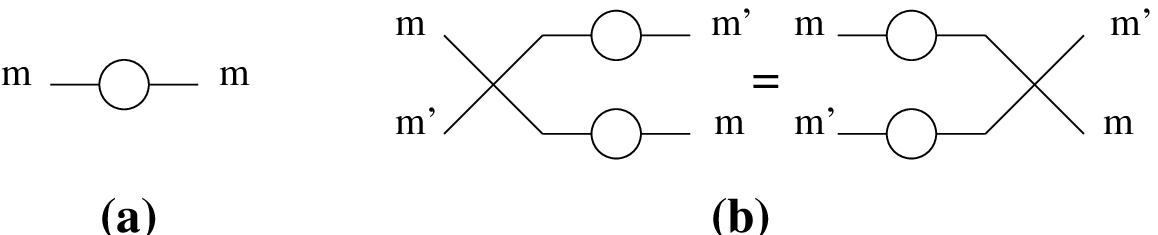}}
\caption{(a) Sklyanin's ${\bf K}$-matrix and (b) a graphical depiction
of \Eq{rkcomm}.}
\label{fig:sklyanink}
\vspace{0.5cm}
\end{center}
\end{figure}

In this section, we shall generalize, following
Sklyanin,\cite{sklyanin-1,sklyanin-2,sklyanin-3} 
the formalism described above to the
case when the boundary conditions in the horizontal direction is not strictly
periodic: instead, 
the $1$'st and $N+1$'st horizontal bonds are to be identified only
up to a ``twist'', implemented using a 
(fixed) linear transformation. We will show later on that the DBCS
model (and also some of its possible generalizations) is some special limiting
case of the inhomogeneus XXX (XXZ) model with such a twisted boundary
conditions.  Consider a diagonal $2\times 2$ matrix ${\bf K_m}$, first
introduced by Sklyanin \cite{sklyanin-1,sklyanin-2}, acting on the horizontal
space $U_m$: \beq {\bf K_m}=\left(\begin{array}{cc}(K_m)_{11}&0 \\0&
    (K_m)_{22}
 \end{array}\right)
\label{kmat}
\eeq
It is easy to check that the following relation holds\cite{generalK}
(illustrated in Fig.~\ref{fig:sklyanink}),
\beq
\left[{\bf \tilde R_{m m'}} , {\bf K_m}{\bf K_{m'}}\right]=0, 
\label{rkcomm}
\eeq
\noindent
where, in accord with our earlier conventions, the subscripts $(m,m')$ specify
the horizontal spaces on which the operators act non trivially.  Let us define
a modified Monodromy matrix as 
\beq 
\tilde{{\cal T}}_m \equiv 
\tilde{{\cal T}} (\l_m) \equiv {\bf K_m}
{\cal T} (\l_m) , \eeq or, in the $2\times 2$ block form of 
Eq.~(\ref{mmat2x2}), 
\beq 
\tilde{{\cal T}_m} \equiv \left( \begin{array}{c c}
    \tilde A_m&\tilde B_m \\ \tilde C_m &\tilde D_m \end{array}\right)= 
\left(
  \begin{array}{c c} (K_m)_{11}&0\\0&(K_m)_{22} \end{array}\right) \left(
  \begin{array}{c c} A_m &B_m \\ C_m & D_m \end{array}\right) \; . 
\eeq 
Eq.~(\ref{rkcomm}) ensures that the new Monodromy matrix 
$\tilde{{\cal T}}$ obeys exactly the same exchange relation 
(\ref{mmexr}) as ${\cal T}$, so that in particular (\ref{tmcomm}) 
and (\ref{commrel}) remain valid also after the substitution 
${\bf T} \rightarrow \tilde {{\bf T}}$, $A \rightarrow \tilde A$,
$B \rightarrow \tilde B$, $C \rightarrow \tilde C$, 
$D \rightarrow \tilde D$. Furthermore, the analogues of
Eqs.~(\ref{rstate})
take the form:
\beqa
\tilde{C}(\l) \ket &=&0,  \nonumber \\
\tilde{A}(\l) \ket &=& K_{11}\ket ,  \nonumber \\
\tilde{D}(\l) \ket &=& K_{22}\prod^N_{i=1} b(\l , \xi_i)\ket .
\label{newrstate}
\eeqa    
It follows that the Bethe vector [cf. \Eq{betv}]
\beq
| \mu_1 , \ldots , \mu^\pdag_P \rangle^{\pdag}_{\bf K}
\equiv \tilde B(\mu_1) \ldots \tilde B(\mu^\pdag_P) \ket 
\label{betv-new}
\eeq
is an eigenstate of 
$\tilde{{\bf T}}(\l)$, 
\beq
\tilde{{\bf T}}(\l) |\mu_1 , \ldots , \mu^\pdag_P \rangle^{\pdag}_{\bf K} = 
\tilde{t}(\l ;\mu_1 , \ldots , \mu^\pdag_P ) 
|\mu_1 , \ldots , \mu^\pdag_P \rangle^{\pdag}_{\bf K} \; ,  
\eeq
with eigenvalue
\beq
\tilde{t}(\l ;\mu_1 , \ldots , \mu^\pdag_P )=
K_{11} \prod^P_{l=1} {1 \over b(\mu_l, \l)} + 
K_{22} \prod^P_{l=1} {1 \over b ( \l, \mu_l )}
\prod^N_{i=1} b(\l , \xi_i)  \; , 
\label{neweigenv} 
\eeq
provided  that the parameters $\mu_j$ satisfy the
following Bethe equations, for $l = 1, \dots, P$:
\beq
K_{11} 
\prod_{\stackrel{l'=1}{l' \ne l}}^P
\frac{b(\mu_l, \mu_{l'})}{b(\mu_{l'}, \mu_l)} = 
K_{22}\prod_{i=1}^N b(\mu_l, \xi_i) \;. 
\label{newbaeq}
\eeq

\section{The ``quasiclassical'' limit }
\label{sec:quasiclassical} 

In this section we show how the DBCS pairing model, or a
generalisation thereof, can be recovered by taking the so-called
``quasi-classical'' limit ($\eta \rightarrow 0 $) of the TIXXX or TIXXZ model,
respectively. We shall present explicitly calculations for the TIXXZ case,
i.e.\ for $\phi(x) = \sinh (x)$; to
recover the corresponding results for the TIXXX case, one simply has to
replace all hyperbolic functions by the corresponding rational ones.

\subsection{Generator for Conserved Operators}
\label{sec:conserved-operators}

Before taking the limit $\eta \rightarrow 0$,
it is convenient to write the inhomogeneity parameters as
\beq
\xi_i=2\varepsilon_i + \eta \; ,
\label{shiftinhom}
\eeq
where the new parameters $\varepsilon_i$'s are
taken to be independent of $\eta$,
and rescale the $\Rmi$
operators by a scalar factor, as follows:
\begin{eqnarray}
\label{eq:rescale}
\Rmi \to  { 2  \Rmi, \over b(\l_m , \xi_i)+1}=
\frac{\sinh (\l_m - 2\varepsilon_i + \eta)}
{\sinh (\l_m - 2\varepsilon_i)
\cosh (\eta)}\Rmi  . 
\end{eqnarray}
These transformations are convenient because, first, then the leading term in
$\Rmi$ is simply a direct product of unit matrices (see \Eq{rmatnew} below);
and second, as we will see later, then many equations transform simply under $
\eta \to - \eta$ (being either symmetric or antisymmetric), which considerably
simplifies all expansions in powers of $\eta$.  When written in terms of a
direct product of $2 \times 2$ Pauli matrices as in Eq.~(\ref{rmatpauli}), the
rescaled $\Rmi$ of (\Eq{eq:rescale}) takes the form \beq \Rmi ={\bf I_m}
\otimes {\bf I_i}+\frac{\tanh \eta}{\tanh (\lllmi)} \bm{\sigma}_{\bf m}^z
\otimes \bm{\sigma}_{\bf i}^z + \frac{2\sinh \eta}{\sinh (\lllmi)} \left(
  \bm{\sigma}_{\bf m}^+ \otimes \bm{\sigma}_{\bf i}^- + \bm{\sigma}_{\bf m}^-
  \otimes \bm{\sigma}_{\bf i}^+ \right).
\label{rmatnew}  
\eeq
Now choose the following form for the {\bf K}-matrix:
\beq
{\bf K}={\bf I}+\frac{\eta}{g} \bm{\sigma}^z ,
\label{kmatnew}
\eeq
and  expand the transfer 
matrix in powers of the parameter $\eta$,
using (\ref{rmatnew}) and (\ref{kmatnew}). This readily yields
\beqa
{\bf \tilde T}(\l_m) \equiv {\rm Tr}_m \{ {\bf K_m R_{mN}\ldots R_{m1}} \} =
2{\bf I}+ \frac{4}{g}\eta^2 {\bf P}(\l_m) +
{\it O}(\eta^3), 
\label{texp} 
\eeqa
where 
\beq
\nonumber
{\bf P}(\l) & =& {1 \over 2}
\sum_{i=1}^N \frac{\bm{\sigma}_{\bf i}^z}{\tanh (\lllzi)}
\\
& & +
g 
\sum_{\stackrel{i,j}{i<j}}^N
\left( \frac{\bm{\sigma}_{\bf i}^z \bm{\sigma}_{\bf j}^z}
{2 \tanh (\lllzi)\tanh (\lllzj) } +  \frac{\bm{\sigma}_{\bf i}^+ 
\bm{\sigma}_{\bf j}^- + 
\bm{\sigma}_{\bf i}^- \bm{\sigma}_{\bf j}^+}
{\sinh (\lllzi)\sinh (\lllzj)}\right) , 
\label{p}
\eeq
and to the commutativity (\ref{tmcomm}) of transfer matrices for 
different spectral parameters guarantees that
\beq
\left[{\bf P}(\l),{\bf P}(\l')\right]=0.
\label{pcomm}
\eeq
${\bf P}(\lambda)$ can be viewed as the generating operator
for all possible conserved operators of the model.

A convenient way of obtaining a \emph{complete} set
of \emph{commuting} conserved operators,
is to  take the residues of ${\bf P}(\l)$ at the points 
$\l =2\varepsilon_i$ for $i=1 \ldots N$,
\begin{eqnarray}
  \label{eq:construct-conserved-operators}
{\bf H_i}={\rm Res}[ {\bf P} (\l); {\l \to 2\varepsilon_i}] \;
= \; \oint_{{\cal C}_i}\frac{d\l}{2\pi i} {\bf P} (\l) \; ,
\end{eqnarray}
where ${\cal C}_i$ is a small contour in the 
complex $\l$-plane, encircling the point $2\varepsilon_i$.
Explicitly evaluating the residues for the present model, one obtains
\beq
{\bf H_i}= 
{ \bm{\sigma}_{\bf i}^z\over 2}  + g
\sum_{\stackrel{j=1}{j \ne i}}^N
\left( \frac{\bm{\sigma}_{\bf i}^z \bm{\sigma}_{\bf j}^z}
{2\tanh (2\varepsilon_i - 2\varepsilon_j) } + \frac{\bm{\sigma}_{\bf i}^+ 
\bm{\sigma}_{\bf j}^- + 
\bm{\sigma}_{\bf i}^- \bm{\sigma}_{\bf j}^+}
{\sinh (2\varepsilon_i - 2\varepsilon_j)}\right) \;. 
\label{gham}
\eeq
Eq.~(\ref{pcomm}) immediately implies that all of 
these operators commute: 
$\left[{\bf H_i},{\bf H_j} \right]=0$.
Furthermore, it is not difficult to show that the
set of all ${\bf H_i}$ is complete, in the sense that
${\bf P}(\lambda)$ can be expressed purely in terms of these
operators. Indeed, ${\bf P}(\l)$ is a  
rational (matrix-valued) function of the variable 
$z\equiv \exp (2\l)$, which is regular at $z\rightarrow 
\infty$ and has  simple poles at $z\rightarrow \exp ( 4\varepsilon_i)$;
it is thus completely determined in terms of the corresponding 
residues, which are equal  
to ${2 \exp (4 \varepsilon_i)\bf H_i}$, so that we have 
\begin{eqnarray}
\label{eq:PintermsofH}
{\bf P}(\lambda) = {\bf P}(\infty)+\sum_{i=1}^N {2 e^{4 \varepsilon_i}
{\bf H_i}  
\over e^{2\lambda} - e^{4 \varepsilon_i}}
\; . 
\end{eqnarray}
The term ${\bf P}(\infty)$ itself also can be expressed via 
${\bf H_i}$: 
\beq
{\bf P}(\infty)=g\left(\sum_{i=1}^N {\bf H_i}\right)^2 +
\sum_{i=1}^N {\bf H_i}-\frac{Ng}{4},
\label{pinfty}
\eeq 
where we have used the fact that
\beq
\sum_{i=1}^N {\bf H_i}=\frac{1}{2}\sum_{i=1}^N \bm{\sigma}_{\bf i}^z \; .
\eeq
The commuting operators ${\bf H_i}$ are in fact  just the 
so-called generalised
Gaudin Hamiltonians. Moroever, in the limit $g\rightarrow \infty$,
in which ${\bf K} = {\bf I}$ so that one recovers
periodic boundary conditions, the ${\bf H_i}/g$ reduce
to the standard Gaudin Hamiltonians ${\bf {H}}_{\bf i}^{\rm Gaudin}$
of Eq.~(\ref{gaudinham}). 

\subsection{Eigenvectors}
\label{sec:eigenvectors}

To obtain the quasiclassical limit of the Bethe eigenvectors $|\mu_1 , \ldots
, \mu^\pdag_P \rangle^{\pdag}_{\bf K}$ of Eq.~(\ref{betv-new}), we have to
investigate the $\eta \rightarrow 0$ limit of the operators $\tilde B$ which
enter in its definition of (\ref{betv-new}).  Now, recall that the operator
$\tilde B_m = \tilde B (\lambda_m)$ is the $(1,2)$ component (in auxiliary
space $U_m$) of the Monodromy matrix $\tilde {\cal T}_m = {\bf K_m
  R_{mN}\ldots R_{m1}}$, which has the following expansion [using
Eqs.~(\ref{rmatnew}) and (\ref{kmatnew})]: \beqa
\tilde {\cal T} (\l_m)  &  =& {\bf I}+\frac{\eta}{g}\bm{\sigma}_{\bf m}^z + \\
& & +\sum_{i=1}^N \left[\frac{\eta}{\tanh (\lllmi)}\bm{\sigma}_{\bf m}^z
  \bm{\sigma}_{\bf i}^z +\frac{2\eta}{\sinh (\lllmi)}\left(\bm{\sigma}_{\bf
      m}^+ \bm{\sigma}_{\bf i}^- + \bm{\sigma}_{\bf m}^- \bm{\sigma}_{\bf i}^+
  \right) \right] +O(\eta^2) \; .  \nonumber
\label{mmexp}
\eeqa
 In this equation,  the only terms having
non-zero (1,2) components and hence contributing
to $\tilde B (\lambda_m)$ are those 
proportional to $\bm{\sigma}_{\bf m}^+$. Thus
we have 
\beq
\label{eq:Bexplicit}
\tilde B(\l)=2\eta S^-(\l) +O(\eta^2),
\eeq 
where
\beq
S^- (\l)=\sum_{i=1}^N \frac{\bm{\sigma}_{\bf i}^- }{\sinh (\lllzi)}.
\label{s-}
\eeq
We see that the quasiclassical (unnormalized) 
Bethe vector of Eq.~(\ref{betv-new}) takes the form
\beq
|\mu_1 \ldots \mu^\pdag_P \rangle^{\pdag}_{\bf K} = S^- (\mu_1) \ldots S^- 
(\mu^\pdag_P)  \ket ,
\label{qbetv}
\eeq
where, as before, the reference state $\ket $ is defined by 
(\ref{ket}).   

\subsection{Eigenvalues}
\label{sec:eigenvalues}

The eigenvalues $h_i$ of the conserved operators
${\bf H_i}$ can be found from the  eigenvalue $p(\lambda)$ of 
their generator 
${\bf P} (\lambda)$. Since  $(4/g) {\bf P} (\lambda)$ 
is the order-$\eta^2$ coefficient 
of the transfer matrix ${\bf \tilde T}(\lambda)$
[Eq.~(\ref{texp})],
its eigenvalue $(4/g) p(\lambda)$ is given by
the order-$\eta^2$ coefficient
of the corresponding eigenvalue $\tilde t(\lambda)$,
which can be found by multiplying \Eq{neweigenv} by the factor
 \beq
\prod_{i=1}^N \frac{\sinh (\l - 2 \varepsilon_i +\eta)}
{\sinh (\l - 2 \varepsilon_i)\cosh (\eta)} 
\eeq
[cf.~(\ref{eq:rescale})] and setting
$\xi_i =  2\varepsilon_i +\eta$ [cf.~(\ref{shiftinhom})]:
\beqa
\tilde t(\l)=\left[ \left(1+\frac{\eta}{g}\right) \prod_{i=1}^N 
\frac{\sinh (\lllzi+\eta)}{ \sinh (\lllzi)\cosh (\eta)}
\prod_{l=1}^P \frac{\sinh (\l -\mu_l -2\eta)}{\sinh (\l -\mu_l)} 
\right. +\\
\left. \left(1-\frac{\eta}{g}\right) \prod_{i=1}^N 
\frac{\sinh (\lllzi-\eta)}{ \sinh (\lllzi)\cosh (\eta)}
\prod_{l=1}^P \frac{\sinh (\l -\mu_l +2\eta)}{\sinh (\l -\mu_l)}
\right]. \nonumber
\label{modeigenv} 
\eeqa
Expanding this expression in  $\eta$, 
the coefficient of $\eta^2$, multiplied by $g/4$, is found to be:
\beqa
p(\l)=  Pg+\frac{1}{2}\sum_{i=1}^N \frac{1}{\tanh (\lllzi)} -
\sum_{l=1}^P \frac{1}{\tanh (\l -\mu_l)}+\nonumber \\
\frac{1}{2}
\sum_{\stackrel{i,i^\prime}{i<i^\prime}}^N
\frac{g}
{\tanh (\lllzi) \tanh(\lllzip)} + 
\sum_{\stackrel{l,l^\prime}{l<l^\prime}}^P
\frac{2g}
{\tanh (\l -\mu_l)\tanh (\l - \mu_{l'})}  \nonumber \\
-\sum_{i}^{N} \sum_{l=1}^P \frac{g}
{\tanh (\lllzi)\tanh (\l -\mu_l)}. 
\label{peigenv}
\eeqa 
The eigenvalues of the generalized 
Gaudin Hamiltonians ${\bf H_i}$, say $h_i$, can be obtained  by 
taking the residues at the points $\l=2\varepsilon_i$:
\beq
h_i=\frac{1}{2}-\sum_{l=1}^P \frac{g}{\tanh (2\varepsilon_i -\mu_l)}+ 
\frac{1}{2}
\sum_{\stackrel{i^\prime =1}{i^\prime \ne i}}^N
\frac{g}
{\tanh (2\varepsilon_i -2\varepsilon_{i^\prime})}
\; .
\label{heigenv}
\eeq
Finally, since all the ${\bf H_i}$
commute, we can immediately write
down the eigenvalue of any function of these operators.
In particular, the general 
Hamiltonian
\beq
\label{genham}
{\bf H}={\cal P}\left({\bf H_1}, \ldots , {\bf H_N}\right), \eeq where ${\cal
  P}$ is some arbitrary polinomial of its arguments, has eigenvalues ${\cal
  P}(h_1, \dots h_N)$.  For example, the general class of models recently
discussed by Amico, Di Lorenzo and Osterloh \cite{ADLO1} in the context of
superconductivity in small grains, is obtained by considering certain second
order polynomials (i.e.\ quadratic combinations of ${\bf H_i}$'s).

\subsection{Bethe equations}

The quasiclassical Bethe state $|\mu_1, \dots , \mu_P \rangle^{\pdag}_{\bf K}$
 is an eigenstate of the generator ${\bf P}(\l)$, and consequently also of
each of the generalized Gaudin Hamiltonians ${\bf H_i}$, only if the
parameters $\mu_l$ satisfy the limit $\eta \rightarrow 0$ of the Bethe
equations (\ref{newbaeq}). The latter are of course not affected by the
rescaling transformation (\ref{eq:rescale}), and take the following form upon
inserting $\xi_i=2\varepsilon_i + \eta $ of (\ref{shiftinhom}): \beqa
\left(1+\frac{\eta}{g}\right) \prod_{\stackrel{l'=1}{l' \ne l}}^P {\sinh
  (\mu_{l'} -\mu_l +2\eta) \over \sinh (\mu_{l'} -\mu_l -2\eta) } =
\left(1-\frac{\eta}{g}\right) \prod_{i=1}^N {\sinh (\mu_l - 2\varepsilon_i
  -\eta) \over \sinh (\mu_{l} -2\varepsilon_i +\eta) } \; .  \nonumber
\label{modbaeq}
\eeqa
In the ``quasiclassical'' limit $\eta \rightarrow 0$ we obtain
the following set of  equations, for $l = 1, \dots, P$,
which may be viewed as generalized Gaudin equations:
\beq
\frac{1}{g}-\sum_{i=1}^N \frac{1}{\tanh (2\varepsilon_i-\mu_l)} + 
\sum_{\stackrel{l'=1}{l' \ne l}}^P
\frac{2}{\tanh (\mu_{l'} -\mu_l)}=0.
\label{geq}
\eeq 
We would like to emphasize that these are the \emph{on-shell}
Bethe equations of quasiclassical limit of the TIXXZ model. In
contrast, in Refs.~\onlinecite{AFF} and \onlinecite{ADLO1}, who did
not consider twisted boundary conditions as we do here, these
equations are \emph{off-shell} Bethe Ansatz equations.

Of course, Eqs.~(\ref{geq}) can be derived, if desired, without reference to
the ABA, by pursuing the following strategy (described in detail in Appendix
B of Ref.~\onlinecite{vDR}): in order to show that the state $|\mu_1 \ldots
\mu^\pdag_P \rangle^{\pdag}_{\bf K}$ 
of Eq.~(\ref{qbetv}) is an eigenstate of any ${\bf H_i}$,
one would commute ${\bf H_i}$ past all the operators $S^- (\mu_l)$ in
Eq.~(\ref{qbetv}) [whose form (\ref{s-}) is reminiscent of the
operators $B_\mu$ defined Ref.~\onlinecite{vDR} if we identify
$\bm{\sigma}_{\bf i}^-$ with $b^\dagger_i$]; this would generate ``unwanted''
terms that only vanish if Eqs.~(\ref{geq}) are satisfied.

\subsection{Specialization to Richardson's equations}

It is straightforward to recover the DBCS model
and Richardson's solution thereof, as summarized in
Section~\ref{Richardson-Gaudin}, by considering
the case $\phi(x) = x$ appropriate for the 
XXX model (instead of the XXZ case $\phi = \sinh x$),
and  replacing everywhere
\begin{eqnarray}
  \label{eq:tanhtox}
  \tanh x \to x, \qquad \sinh x \to x \; . 
\end{eqnarray}
First, we note that the generalized Gaudin equations
(\ref{geq}) then reduce to Richardson's equations
(\ref{eq:richardson-eigenvalues-cc}). Furthermore, the generalized
Gaudin Hamiltonians ${\bf H_i}$ of (\ref{gham}) reduce to the
form given in Eq.~(\ref{eq:constants-of-motion-DBCS}) for
the conserved operators of the DBCS model. This fact
was noted by Sklyanin himself in a side remark in
Ref.~\onlinecite{sklyanin-2}, and first derived by him already
in 1989 in Ref.~\onlinecite{sklyanin-3}. However, he was
at the time unaware of the fact that the resulting ${\bf H}_i$ 
were useful in the context of the DBCS model, and in particular,
that they 
be used to construct the Hamiltonian $H_U$
of (\ref{eq:HUspin-variables}) of the DBCS model.
It it is straightforward to check that this can
be done through the following construction:
\beq
H_U ({\bf H_i}) =\sum_{i=1}^N \Bigl[ \left(g-2\varepsilon_i\right){\bf H_i}+
\left(\varepsilon_i- 3g/4 \right) \Bigr] \; + \; 
g\Bigl (\sum_{i=1}^N {\bf H_i}\Bigr)^2 \; . 
\label{hu}
\eeq
To calculate its eigenvalues ${\cal E}_P = H_U (h_i)$
explicitly, the following identities [derived by repeated use of
 (\ref{heigenv}) and (\ref{geq})] are useful:
\begin{mathletters}
\label{eq:trivialidenties}
\beq
\sum_{i=1}^N h_i & = & \frac{N}{2}-\sum_{l=1}^P\sum_{i=1}^N
\frac{g}{2\varepsilon_i-\mu_l}=\frac{N}{2}-
\sum_{l=1}^P\left(1+
\sum_{\stackrel{l'=1}{l' \ne l}}^P
\frac{2g}{\mu_{l'}-\mu_l}\right)
=\frac{N}{2}-P,
\label{energycal1}
\\
\label{energycal2-a}
\sum_{i=1}^N 2\varepsilon_ih_i &  = & \sum_{i=1}^N \varepsilon_i
+ \sum_{\stackrel{i,i^\prime}{i \ne i^\prime}}^N
\frac{g\varepsilon_i}
{2\left(\varepsilon_i-\varepsilon_{i'}\right)} 
- \sum_{i=1}^N\sum_{l=1}^P \frac{2g\varepsilon_i}{2\varepsilon_i-\mu_l}
\\
\label{energycal2}
& = &  \sum_{i=1}^N \varepsilon_i 
+ g N (N- 1)/4 
-  \Bigl[ \sum_{l=1}^P \mu_l + g P N - g P (P -1) \Bigr]
\; , 
\eeq
where the last term of (\ref{energycal2}) was obtained
from the last term of (\ref{energycal2-a}) by rewriting the latter as follows:
\beq
\sum_{i=1}^N\sum_{l=1}^P \frac{g(2\varepsilon_i-\mu_l+
\mu_l)}{2\varepsilon_i-\mu_l}=gPN+\sum_{l=1}^P 
\mu_l\Bigl (  1+
\sum_{\stackrel{l'=1}{l' \ne l}}^P
\frac{2g}{\mu_{l'}-\mu_l}\Bigr)  \; . 
\label{energycal3}
\eeq
\end{mathletters}
Using Eqs.~(\ref{energycal1}) and (\ref{energycal2}),
it is straightforward to check that ${\cal E}_P$
reduces to the simple form ${\cal E}_P = \sum_{l=1}^P \mu_l$ of 
Eq.~(\ref{eq:truebosoneigenvalues}).

\section{Conclusions}
\label{sec:conclusions}

The TIXXZ results of the previous section for the conserved operators
${\bf H_i}$, their eigenvalues $h_i$ and eigenvectors $|\mu_1 , \ldots
, \mu^\pdag_P \rangle^{\pdag}_{\bf K} $, and the corresponding
consistency condition (\ref{geq}), have been found independently
before by Amico, Di Lorenzo and Osterloh \cite{ADLO1}. They managed to
construct the ${\bf H_i}$ apparently by inspection, without presenting
a systematic approach for their derivation, and in their approach the
consistency condition (\ref{geq}) appears as a set of off-shell Bethe
Ansatz equations. In our work, we presented a systematic derivation of
these results from a vertex model with twisted boundary conditions,
and the consistency condition (\ref{geq}) corresponds directly to the
\emph{on-shell} Bethe Ansatz equations of this model. Thus, we hope to have
have shed some additional light on the reasons why the DBCS model and
its generalizations are integrable and Bethe-Ansatz solvable, and on
the underlying algebraic structure of the solutions. We hope that our
works shows the way towards further progress in applying the powerful
formalism of the ABA to the DBCS and related models, e.g.\ for the
calculation of correlation functions\cite{AO2001} such as $\langle
S^z_i S^z_j \rangle$ or $\langle S^-_i S^+_j \rangle$, which are of
importance for understanding the nature of pairing correlations in
nanoscale superconducting grains.\cite{vD,vDR}

\bigskip

\emph{Acknowledgements:---} We acknowledge very fruitful discussions with and
stimulating encouragement from R. Flume and V. Rittenberg,
and helpful comments from L. Amico.  RP has been
supported in part by the INTAS grant 99-1459 and the A. v. Humboldt foundation
of the Federal Republic of Germany.

\end{document}